# Simultaneous determination of neutron-induced fission and radiative-capture cross sections from decay probabilities obtained with a surrogate reaction

R. Pérez Sánchez[1,2], B. Jurado[1,*], V. Méot[2], O. Roig[2], M. Dupuis[2], O. Bouland[3], D. Denis-Petit[1,2], P. Marini[2], L. Mathieu[1], I. Tsekhanovich[1], M. Aïche[1], L. Audouin[4], C. Cannes[4], S. Czajkowski[1], S. Delpech[4], A. Görgen[5], M. Guttormsen[5], A. Henriques[1], G. Kessedjian[6], K. Nishio[7], D. Ramos[4], S. Siem[5] and F. Zeiser[5]

[1]CENBG, CNRS/IN2P3-Université de Bordeaux, Chemin du Solarium, B.P. 120, F-33175 Gradignan, France
[2]CEA, DAM, DIF, F-91297 Arpajon, France
[3]CEA-Cadarache, DEN/DER/SPRC/LEPh, F-13108 Saint Paul lez Durance, France
[4]Université Paris-Saclay, CNRS/IN2P3, IJC Lab, 91405 Orsay, France
[5]Department of Physics, University of Oslo, 0316 Oslo, Norway
[6]Université Grenoble-Alpes, Grenoble INP, CNRS, LPSC-IN2P3, 38000 Grenoble, France
[7]JAEA, Tokai, Ibaraki 319-1195, Japan

**Abstract**: Reliable neutron-induced reaction cross sections of unstable nuclei are essential for nuclear astrophysics and applications but their direct measurement is often impossible. The surrogate-reaction method is one of the most promising alternatives to access these cross sections. In this work, we successfully applied the surrogate-reaction method to infer for the first time both the neutron-induced fission and radiative-capture cross sections of $^{239}$Pu in a consistent manner from a single measurement. This was achieved by combining simultaneously-measured fission and γ-emission probabilities for the $^{240}$Pu($^{4}$He, $^{4}$He') surrogate reaction with a calculation of the angular-momentum and parity distributions populated in this reaction. While other experiments measure the probabilities for some selected γ-ray transitions, we measure the γ-emission probability. This enlarges the applicability of the surrogate-reaction method.

Chemical elements heavier than iron are produced in stars mainly via neutron-induced reactions in the slow (*s*) and rapid (*r*) neutron-capture processes [1]. The recent detection of gravitational waves from the merger of two neutron stars [2], and the subsequent kilonova, consistent with being powered by the radioactive decay of nuclei synthesized by the r-process [3, 4], demonstrated that neutron-star mergers are an important r-process site. However, it is not yet clear if the r-process abundance distribution in the solar system is the result of one or multiple scenarios. The measurement of neutron-induced cross sections of key neutron-rich nuclei is an essential component to answer this question [5]. In the neutron-star merger scenario, heavy nuclei are produced. Their fission can strongly influence r-process observables such as abundance patterns and light curves [5, 6, 7]. Key physical quantities for understanding the impact of fission are neutron-induced fission cross sections and fission barriers. Neutron-induced reaction cross sections of radioactive nuclei are also crucial ingredients for the simulation of advanced nuclear energy systems [8] and the production of diagnostic or therapeutic radionuclides [9, 10].

---

[*] jurado@cenbg.in2p3.fr

The direct measurement of neutron-induced cross sections of unstable nuclei is very complicated due to the radioactivity of the targets involved. Measurements in inverse kinematics are not possible because free-neutron targets are not available. On the other hand, the different theoretical model predictions of these cross sections strongly diverge, reaching variations up to several orders of magnitude, due to the difficulties in describing the de-excitation process of the nucleus formed by neutron absorption. Indeed, the de-excitation process is ruled by fundamental properties (nuclear level densities, fission barriers, etc.) for which the existing nuclear models give very different results when no experimental data are available [11, 12].

Several indirect approaches have been developed to overcome these issues [13, 14], such as the Oslo [14] and the surrogate-reaction [15] methods. While all these techniques are used to determine radiative neutron-capture cross sections, only the surrogate-reaction method has also been applied to infer fission cross sections. In the surrogate reaction method [15], the excited nucleus produced in the neutron-induced reaction of interest is formed via an alternative, experimentally-accessible charged-particle-induced reaction. The decay probabilities obtained with the alternative (surrogate) reaction can be used to constrain the models that describe the de-excitation process. In the Hauser-Feshbach (HF) statistical reaction formalism [16], the decay probabilities $P_{i,\chi}$ obtained with the surrogate ($i=s$) and the neutron-induced ($i=n$) reactions are given by:

$$P_{i,\chi}(E^*) = \sum_{J^\pi} F_i(E^*,J^\pi) \cdot G_\chi(E^*,J^\pi) \tag{1}$$

where $F_i(E^*,J^\pi)$ are the probabilities to form the excited nucleus in a state of spin $J$ and parity $\pi$ at an excitation energy $E^*$ by the corresponding reaction, and $G_\chi(E^*,J^\pi)$ are the probabilities that the nucleus decays from that state via decay channel $\chi$. The factorized form in eq. (1) reflects the essential assumption of the HF model that a compound nucleus (CN) is formed whose decay is independent of the way it was formed [17]. At a sufficiently high $E^*$ the Weisskopf-Ewing (WE) limit applies, i.e. the probabilities $G_\chi$ become independent of $J^\pi$ [15], thus $P_{s,\chi} \approx P_{n,\chi}$ and one may infer the neutron-induced cross section $\sigma_{n,\chi}$ by applying:

$$\sigma_{n,\chi}(E_n) \cong \sigma_{CN}(E_n) \cdot P_{s,\chi}(E^*) \tag{2}$$

where $\sigma_{CN}(E_n)$ is the cross section for the formation of a CN after the absorption of a neutron of energy $E_n$. $\sigma_{CN}$ can be calculated with the optical model with an uncertainty of about 5 % for nuclei near the stability valley [15]. $E_n$ and $E^*$ are related via:

$$E_n = \frac{A+1}{A} \cdot (E^* - S_n) \tag{3}$$

where $S_n$ is the neutron separation energy of the CN and $A$ is the mass number of the target nucleus in the neutron-induced reaction. Surrogate reactions were first used to infer neutron-induced fission cross sections $\sigma_{n,f}$ [18]. Several measurements, e.g [19], have shown that the $\sigma_{n,f}$ obtained with eq. (2) (i.e. by multiplying the measured fission probabilities $P_{s,f}$ by $\sigma_{CN}$), are in good agreement with directly-measured neutron data. However, the WE approximation fails when applied to infer $\sigma_{n,f}$ of even-even fissioning nuclei [20, 21] and neutron-induced radiative-capture cross sections $\sigma_{n,\gamma}$ [22, 23, 24]. The observed disagreement has been attributed to the differences between the probability distributions $F_i$ populated in the surrogate and neutron-induced reactions and the failure of the WE limit [15]. Recently, Escher et al. [25] and Ratkiewicz et al. [26]

have shown that the probabilities for observing specific γ-ray transitions in surrogate reactions combined with the calculated distributions $F_s$ can be used to tune HF model parameters and significantly improve the predictions for $\sigma_{n,\gamma}$ of $A \approx 90$ nuclei.

In this work we make an important step forward and apply the surrogate-reaction method to *simultaneously* infer $\sigma_{n,f}$ and $\sigma_{n,\gamma}$ of an even-even actinide nucleus. In particular, we have used the $^{240}$Pu($^4$He,$^4$He')$^{240}$Pu* reaction as a surrogate for the n+$^{239}$Pu reaction. We also stress a significant difference with respect to [25] and [26], namely that, instead of measuring the probabilities for some selected γ-ray transitions, we measure the γ-emission probability $P_{s,\gamma}$, i.e. the probability that the CN releases its entire $E^*$ by emitting a cascade of γ rays. Using $P_{s,\gamma}$ can lead to a more reliable determination of the HF model parameters since it does not involve modeling the details of the γ-ray cascade, which requires including the branching ratios of many low-lying transitions that are often unknown. This is however much less critical for an even-even CN, as considered in [26], because there is generally a strong-collecting $2^+ \rightarrow 0^+$ γ-ray transition. Yet, measuring the $P_{s,\gamma}$ of fissionable nuclei is very challenging as it requires removing the intense background of γ rays emitted by the fission fragments.

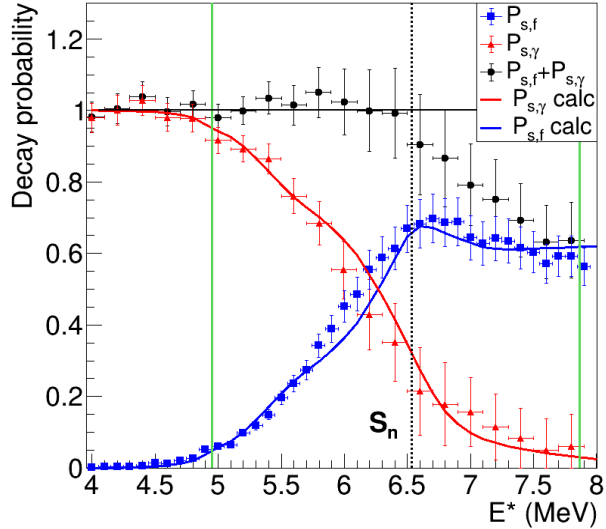

**Figure 1:** Decay probabilities for fission (blue squares) and γ emission (red triangles) measured for the $^{240}$Pu($^4$He,$^4$He')$^{240}$Pu* reaction as a function of the excitation energy $E^*$ of $^{240}$Pu*. The sum of the two probabilities is given by the black circles. The $P_{s,f}$ and $P_{s,\gamma}$ calculated with eq. (1) and the values of the adjusted Talys parameters that led to the minimum $\chi^2$ value are represented by the blue and red solid lines, respectively. The $E^*$ range used for parameter adjustment is delimited by the vertical green lines. The vertical dotted line indicates the neutron separation energy $S_n$ of $^{240}$Pu. The horizontal black line at a constant value of 1 serves to guide the eye.

The details of our experiment, and the data and uncertainty analysis can be found in [27], here we only describe the main features. A 30 MeV beam of $^4$He delivered by the Tandem accelerator of the ALTO facility in Orsay (France) impinged on a 100 μg/cm$^2$ PuO$_2$ target deposited on a carbon support of the same areal density. The $P_{s,f}$ ($P_{s,\gamma}$) was obtained from the measured number of scattered $^4$He' and of fission fragments (γ-ray cascades) detected in coincidence with the $^4$He'. The $^4$He' nuclei were detected with two position-sensitive silicon telescopes centered at a polar angle $\theta$ of 138.5° with respect to the beam axis. Fission fragments were detected with an array of solar cells and γ rays with four C$_6$D$_6$ liquid scintillators and five high-purity germanium detectors. We removed the γ rays emitted by $^{239}$Pu by applying a threshold, which varied gradually from 270 keV to 1.5 MeV with increasing $E^*$. The number of γ cascades emitted by the fission fragments was determined by calculating the ratio between the number of triple coincidences $^4$He'/fission-fragment/γ-ray-cascade and the fission efficiency. To determine the latter number of cascades with sufficient precision, it was necessary to maximize all the detection efficiencies and at the same time determine the fission and the γ-cascade detection efficiencies with a relative uncertainty below 10 and 20 %, respectively, which is a difficult task, see [27]. The decay probabilities were measured at seven $^4$He' scattering angles $\theta_{4He'}$, ranging from 120.7 to 156.0° in steps of about 6°, and no significant differences were observed. Therefore, we determined the weighted average of the probabilities obtained at the individual $\theta_{4He'}$ angles, the results are shown in Fig. 1. The error bars include statistical and systematic uncertainties, as well as the covariance between the measured quantities. The $E^*$ resolution was 100 keV.

It is the first time that both $P_{s,f}$ and $P_{s,\gamma}$ of $^{240}$Pu* have been simultaneously measured. This provides a stringent test of the used experimental method. Fig. 1 illustrates that below 4.5 MeV only the γ-decay channel is open and the associated probability is 1. Between 4.5 MeV and $S_n$ fission and γ emission are the only open decay channels and the sum of their probabilities must be 1. Our data satisfy this condition within the uncertainty limits of the measurement, see the black dots in Fig. 1. This validates our experimental procedure. For $E^* > S_n$ the sum of both probabilities is no longer 1 as neutron emission becomes possible and competes with fission and γ emission. The emission of charged particles is massively hindered by the Coulomb barrier.

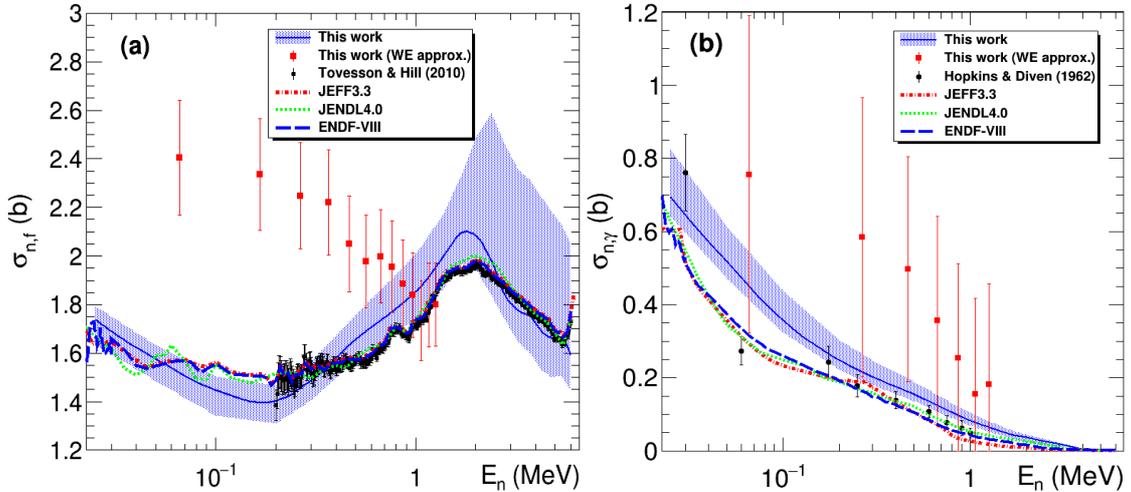

**Figure 2:** Neutron-induced fission (a) and radiative-capture (b) cross sections of $^{239}$Pu as a function of neutron energy. The red squares are the cross sections obtained with the WE approximation (eq. 2). The cross sections calculated with the parameters deduced from the measured decay probabilities are shown as blue solid lines. The shaded blue areas indicate the associated uncertainties. The dash-dotted, dotted and dashed lines represent different evaluations. The black dots indicate the neutron-induced data of [32] (a) and [33] (b).

We first assumed the validity of the WE approximation and derived $\sigma_{n,f}$ and $\sigma_{n,\gamma}$ using eq. (2). More precisely, we converted the $E^*$ into equivalent neutron energy $E_n$ with eq. (3) and multiplied the measured probabilities for fission $P_{s,f}(E_n)$ and γ emission $P_{s,\gamma}(E_n)$ by $\sigma_{CN}(E_n)$. $\sigma_{CN}(E_n)$ was calculated with Talys 1.9 [28] using the parameters of the optical model given in JENDL-4.0 [29]. When $E^* < S_n$, the equivalent $E_n$ is negative. Therefore, only the data within the range $S_n < E^* < 7.8$ MeV are considered, see red squares in Fig. 2. The resulting cross sections are compared in Fig. 2 with the evaluations JEFF-3.3 [30], JENDL-4.0 [29] and ENDF/B-VIII.0 [31], the most recent measurement of $\sigma_{n,f}$ [32] and the only data for $\sigma_{n,\gamma}$, which to our knowledge extend to $E_n$=1 MeV [33]. We observe a clear disagreement for fission, especially at low energies. For $\sigma_{n,\gamma}$, the red squares are well above the evaluations and neutron-induced data. However, due to the large uncertainties, no clear conclusion can be drawn.

To go beyond the WE hypothesis, we calculated the $F_S(E^*,J^\pi)$ distribution populated by the $^{240}$Pu($^4$He,$^4$He') reaction. A lane-consistent parameterization of the Jeukenne-Leujenne-Mahaux central nucleon-nucleon (NN) effective interaction [34] was employed within the double-folding framework following [35]. The excited states of $^{240}$Pu were described as members of rotational bands built from intrinsic excitations of the axially-deformed ground state. These excitations were determined within the quasi-particle random-phase approximation [36]. Inelastic cross sections to each of these states were determined through the coupled-channel framework using a simplified coupling scheme. This method has been successfully applied to determine neutron inelastic cross sections and the corresponding $J^\pi$ distributions of actinides [37]. The extension to $^4$He scattering within the double-folding framework was benchmarked considering elastic and inelastic cross sections for spherical and deformed nuclei [38]. Our results for $F_S(E^*,J^\pi)$ do not noticeably change across the angular range covered by our experiment, which is consistent with the observed insensitivity of the measured decay probabilities to $\theta_{4He'}$. The calculated $F_s$ distribution at $E^*$=7.5 MeV and $\theta_{4He'}$= 140° is shown in Fig. 3. The average CN spin $\langle J \rangle$ obtained for the $^{240}$Pu($^4$He,$^4$He') reaction is 5.6 ℏ. This value is significantly larger than the $\langle J \rangle$ populated by the absorption of a neutron leading to $E^*$=7.5 MeV, which according to Talys 1.9 is $\langle J \rangle \approx$1ℏ.

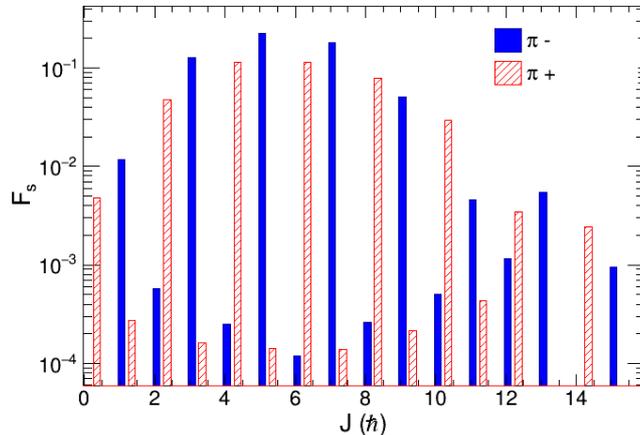

**Figure 3:** Calculated spin-parity distribution $F_s$ of $^{240}$Pu$^*$ populated by the $^{240}$Pu($^4$He,$^4$He') reaction at $E^*$ =7.5 MeV and $\theta_{4He'}$ =140°.

We then used eq. (1) to determine $P_{s,f}$ and $P_{s,\gamma}$ with the calculated $F_S(E^*,J^\pi)$ and the probabilities $G_\chi(E^*,J^\pi)$ calculated with the HF formalism of Talys 1.9 [28]. The values and the uncertainties of several key Talys parameters needed to model the decay of $^{240}$Pu* were tuned to reproduce the experimental $P_{s,f}$ and $P_{s,\gamma}$ in the range $E^* \in$ [5,7.8] MeV with a Bayesian analysis [39]. We assumed Gaussian prior distributions for the parameter values, as discussed below. Finally, these parameters were implemented in Talys 1.9 to calculate $\sigma_{n,f}$ and $\sigma_{n,\gamma}$. The calculation of the neutron-induced cross sections considered width-fluctuation corrections as described in [28].

The following main ingredients were used to model the de-excitation of $^{240}$Pu*. The level densities above the ground states of $^{239}$Pu and $^{240}$Pu were described with the Gilbert-Cameron formula using the recommended parameters of RIPL 3 [40]. The γ-ray Strength Function (γSF) was determined with the model of Goriely [41]. We assumed a one-dimensional, double-humped fission barrier. The headband transition states on top of the fission barriers and the states in the well that separates the two barriers (the Class II states) were initially taken from [42], but to reproduce our data only the $0^+$, $0^-$, $1^-$ and $2^+$ Class II states were retained. In addition, seven other parameters were adjusted to our data: a normalization factor for the γSF [43], the heights and the widths of the two fission barriers, and the two temperature parameters of the Gilbert-Cameron formula at the fission-barrier deformations. The transition and Class II states, and the seven free parameters are the quantities to which $\sigma_{n,f}$ and $\sigma_{n,\gamma}$ are most sensitive.

By individually varying the seven free parameters and comparing the resulting probabilities ($P_{s,f}$ and $P_{s,\gamma}$) to the experimental data shown in Fig. 1, we made an initial coarse estimation of the parameter values and defined the parameter confidence intervals (CI) for a confidence level (CL) of 68%. Afterwards, a random sampling of all the parameter values was done assuming that they followed Gaussian distributions with a standard deviation equal to the CI. This led to thousands of calculations of $P_{s,f}$ and $P_{s,\gamma}$. A $\chi^2$ test was done (CL= 95%) that led to the selection of an ensemble including about 5% of the different sets of parameter combinations. The combination of parameters leading to the $P_{s,f}$ and $P_{s,\gamma}$ that minimized the $\chi^2$ (see blue and red solid lines in Fig. 1) was used to calculate $\sigma_{n,f}$ and $\sigma_{n,\gamma}$, represented by the blue solid curves in Fig. 2. The associated uncertainties correspond to two times the root mean squared (RMS) deviation of the calculations obtained with the retained sets of parameter combinations. With the chosen uncertainties for $\sigma_{n,f}$ and $\sigma_{n,\gamma}$ of 2·RMS our adjustment procedure converged and led to a first fission-barrier height of 5.98 ± 0.02 MeV and a second barrier height of 5.00 ± 0.23 MeV. The widths of the first and second fission barriers are 0.83 ± 0.03 MeV and 0.7 ± 0.1 MeV, respectively. These values are very close to literature [40, 44, 45]. In addition, we obtained a normalization factor for the γSF of 2.8 ± 0.3, which, combined with the model of [41], gives a γSF for $^{240}$Pu that compares well with the experimental results of [46]. The good agreement of these model ingredients with previous work demonstrates the validity of our adjustment procedure.

The relative uncertainties of the calculated cross sections vary from 8% at $E_n$=100 keV to a maximum of 20% at $E_n$=5 MeV. They are smaller than the uncertainties obtained with the WE approximation. The reason is that all the free parameters, which determine both $\sigma_{n,f}$ and $\sigma_{n,\gamma}$, were fixed using experimental points at $E^*$ well below $S_n$, whose uncertainties are significantly smaller than the ones at higher $E^*$. Note that for a fissile nucleus like $^{240}$Pu, whose fission barrier is lower than $S_n$, the range $E^* < S_n$ is not accessible in neutron-induced measurements and our decay probabilities provide unique and precious information to fix the parameters that determine the passage through the fission barrier.

Our result for $\sigma_{n,f}$ agrees rather well with the evaluations and the directly-measured neutron-induced cross sections. We also observe a fairly good agreement for $\sigma_{n,\gamma}$, although our data are somewhat above the evaluations (with the largest discrepancy of about 50% at 60 keV) and most data points of [33]. Such a good level of agreement demonstrates that we have been able to account for the spin/parity differences between the considered surrogate and neutron-induced reactions and obtain, for the first time, reliable results for *both* fission and radiative-capture cross sections. Therefore, the present work shows that surrogate reactions are a powerful tool to fix key model parameters and provide good-quality predictions for neutron-induced reaction cross sections of fissile and/or short-lived nuclei. In the near future, we foresee to apply the same approach to infer the neutron-induced cross sections of other fissile nuclei like $^{242,244}$Pu. However, many unstable nuclei of interest can only be reached by performing experiments in inverse kinematics. Heavy-ion storage rings provide radioactive beams of extraordinary quality and enable the use of ultrathin, windowless targets. Therefore, in the longer term we will combine surrogate reactions and storage rings to simultaneously measure the decay probabilities of all open decay channels for many short-lived nuclei with high precision [47].

**Acknowledgements**

We warmly thank the staff of the tandem accelerator of the ALTO facility at Orsay for their support during the experiment and Pascal Romain for fruitful discussions. This work has been supported by the French "défi interdisciplinaire" NEEDS and by the European Commission within the EURATOM FP7 Framework Program through CHANDA (project n$^o$ 605203).

[1] E. M. Burbidge et al., Rev. Mod. Phys. **29**, 547 (1957).
[2] B. P. Abbott et al., Phys. Rev. Lett. **119**, 161101 (2017).
[3] D. M. Siegel, Eur. Phys. J. A **55**, 203 (2019).
[4] D. Watson et al., Nature **574**, 497 (2019).
[5] T. Kajino et al., Progress in Particle and Nuclear Physics **107**, 109 (2019).
[6] S. Goriely, Eur. Phys. J. A **51**, 22 (2015).
[7] N. Vassh et al., J. Phys. G: Nucl. Part. Phys. **46**, 065202 (2019).
[8] N. Colonna et al., Energy Environ. Sci. **3**, 1910 (2010).
[9] S. M. Qaim et al., IAEA Technical Reports Series No. 473 (2011).
[10] S. M. Qaim, Nuclear Medicine and Biology **44**, 31 (2017).
[11] M. Arnould and S. Goriely, Phys. Reports **384**, 1 (2003).
[12] M. Arnould et al., Phys. Reports **450**, 97 (2007).
[13] J. E. Escher, A. P. Tonchev, J. T. Burke, P. Bedrossian, R. J. Casperson, N. Cooper, R. O. Hughes, P. Humby, R. S. Ilieva, S. Ota et al., Eur. Phys. J. Web Conf. **122**, 12001 (2016).
[14] A.C. Larsen, A. Spyrou, S.N. Liddick, M. Guttormsen, Progress in Particle and Nuclear Physics **107,** 69 (2019).
[15] J. E. Escher, J.T. Burke, F.S. Dietrich, N. D. Scielzo, I. J. Thompson, W. Younes, Rev. Mod. Phys. **84**, 353 (2012).
[16] W. Hauser and H. Feshbach, Phys. Rev. **87**, 366 (1952).
[17] N. Bohr, Nature **137**, 344 (1936).
[18] J. D. Cramer and H.C. Britt, Phy. Rev. C **2**, 6 (1970).
[19] G. Kessedjian, B. Jurado, M. Aïche, G. Barreau, A. Bidaud, S. Czajkowski et al., Phys. Lett. B **692**, 297 (2010).
[20] W. Younes and H. C. Britt, Phys. Rev. C **67**, 024610 (2003).
[21] O. Bouland, Phys. Rev. C **100**, 064611 (2019).
[22] N. D. Scielzo, J. E. Escher, J. M. Allmond, M. S. Basunia, C. W. Beausang, L. A. Bernstein et al, Phys. Rev. C **81**, 034608 (2010).
[23] G. Boutoux, B. Jurado, V. Méot, O. Roig, L. Mathieu, M. Aïche et al., Phys. Lett. B **712**, 319 (2012).

[24] Q. Ducasse, B. Jurado, M. Aïche, P. Marini, L. Mathieu, A. Gorgen, et al., Phys. Rev. C **94,** 024614 (2016).
[25] J. E. Escher, J.T. Burke, R.O. Hughes, N. D. Scielzo, R.J. Casperson, S. Ota, H.I Park, A. Saastamoinen, T. J. Ross, Phys. Rev. Lett. **121**, 052501 (2018).
[26] A. Ratkiewicz, J. A. Cizewski, J.E. Escher, G. Potel, J. T. Burke, R. J. Casperson et al., Phys. Rev. Lett. **122**, 052502 (2019).
[27] R. Pérez Sánchez, B. Jurado, P. Marini, M. Aiche, S. Czajkowski, D. Denis-Petit et al., Nucl. Instrum. Meth. A **933**, 63 (2019).
[28] Talys 1.9 users' manual, 2017. www.talys.eu/documentation.
[29] K. Shibata et al., J. Nucl. Sci. Technol. **48**, 1 (2011).
[30] https://www.oecd-nea.org/dbdata/jeff/jeff33/
[31] D. Brown et al., Nucl. Data Sheets **148**, 1 (2018).
[32] F. Tovesson and T. S. Hill, Nuclear Science and Engineering **165**, 224 (2010).
[33] J. C. Hopkins and B. C. Diven, Nuclear Science and Engineering **12**, 169 (1962).
[34] E. Bauge, J. P. Delaroche and M. Girod, Phys. Rev. C **63**, 024607 (2001).
[35] A. Hogenbirk et al., Phys. Lett. B **223**, 282 (1989).
[36] S. Péru, G. Gosselin, M. Martini, M. Dupuis, S. Hilaire, J. C. Devaux, Phys. Rev. C **83**, 014314 (2011).
[37] M. Dupuis et al., Eur. Phys. J. A **51**, 168 (2015).
[38] M. Dupuis et al., to be published
[39] D.L. Smith, Probability, statistics, and data uncertainties in nuclear science and technology, vol.4, (OECD, Paris, 1991) 269p
[40] R. Capote et al., Nuclear Data Sheets **110**, 3107 (2009).
[41] S. Goriely, Phys. Lett. B **436**, 10 (1998).
[42] J. E. Lynn, P. Talou, and O. Bouland, Phys. Rev. C **97**, 064601 (2018).
[43] J. Kopecky and M. Uhl, Phys. Rev. C **41**, 1941 (1990).
[44] S. Bjoernholm and J. E. Lynn, Rev. Mod. Phys. **52**, 725 (1980).
[45] K.-H. Schmidt, B. Jurado, C. Amouroux and C. Schmitt, Nuclear Data Sheets **131**, 107 (2016).
[46] F. Zeiser, G. M. Tveten, G. Potel, A. C. Larsen, M. Guttormsen, T. A. Laplace et al., Phys. Rev. C **100**, 024305 (2019).
[47] B. Jurado, P. Marini, L. Mathieu, M. Aiche, S. Czajkowski, I. Tsekhanovich et al. EPJ Web of Conferences **146**, 11006 (2017).